\documentclass{svjour3} 
\usepackage{graphics}
\usepackage{amsmath,amssymb}
\usepackage{natbib}
\usepackage{bm}

\smartqed
\journalname{CeMDA}

\def\dm#1{#1}

\begin{document}

\title{Lie-series for orbital elements}
\subtitle{I. The planar case}

\author{Andr\'as P\'al}

\institute{%
A. P\'al 
	\at Konkoly Observatory of the MTA Research Centre for Astronomy and Earth Sciences, Budapest, Hungary and 
	\at Department of Astronomy, Lor\'and E\"otv\"os University, Budapest, Hungary \\    
E-mail: apal@szofi.net}

\maketitle

\begin{abstract}
Lie-integration is one of the most efficient algorithms 
for numerical integration of ordinary 
differential equations if \dm{high precision is needed for longer terms}.
The method is based on the computation of the Taylor-coefficients of the
solution as a set of recurrence relations. 
In this paper we present these recurrence formulae for orbital elements and
other integrals of motion for the planar $N$-body problem.
We show that if the reference frame is fixed to one of the bodies
-- for instance to the Sun in the case of the Solar System --,
the higher order coefficients for all orbital elements and integrals 
of motion depend only on the mutual terms corresponding to the orbiting bodies. 
\end{abstract}

\section{Introduction}
\label{sec:introduction}

Due to the lack of analytical solutions, numerical integration is 
required to solve the equations of motion of the gravitational $N$-body 
problem for almost any initial conditions for $3\le N$. There are many 
textbooks with algorithms related to general purpose numerical integration of
ordinary differential equations 
\citep[ODEs, see e.g.][for an introduction]{press2002}. 
\dm{In principle, if we have to solve the equation 
$\dot x_i=f_i(\mathbf{x})$, where $\mathbf{x}=(x_1,\dots,x_N)$, then
the respective Lie-operator is defined as 
\begin{equation}
L=\sum\limits_{i=1}^Nf_i\frac{\partial}{\partial x_i}.
\end{equation}
The solution of the equation after time $\Delta t$ is then written in the form
\begin{equation}
x(t+\Delta t)=\exp\left(\Delta t\cdot L\right)x(t)=\sum\limits_{k=0}^{\infty}\frac{\Delta t^k}{k!}L^kx(t).
\end{equation}
The finite approximation of the above sum is called Lie-integration
\citep[see also][]{grobner1967}. The higher order derivatives 
can efficiently be computed using recurrence relations where 
the derivatives $L^{k+1}x(t)$ are expressed
as functions of $L^{\ell}x(t)$, where $0\le\ell\le k$.}
The method has many advantages: it is one
of the most efficient methods if we consider long-term and high
precision computations, adaptive forms can be implemented without
losing computation time, roundoff errors are smaller than 
other algorithms, etc. \citep[see e.g.][]{pal2007,hanslmeier1984}.
However, the need of derivations of the respective recurrence 
series for any new problem is a major drawback. 

First, \cite{hanslmeier1984} have obtained the recurrence relations
for the $N$-body problem, taking into account mutual and purely
Newtonian gravitational forces. Soon after, the relations have been 
derived for the restricted three-body problem \citep{delva1984}.
Many methods for stability analysis require the computation of 
linearized equations. The relations for the linearized $N$-body problem
-- including the equations where one of the bodies is fixed --
have been presented by \cite{pal2007}. The algorithm of Lie-integration
has widely been applied for stability studies related to
known planetary systems \citep[see e.g.][]{asghari2004} or special resonant
systems \citep[see e.g.][]{funk2013}. In addition,
more sophisticated semi-numerical methods can be based on the Lie-series
\citep[see e.g.][about the numerical computation
of partial derivatives of coordinates and velocities with respect
to the initial conditions and the direct applications for 
exoplanetary analysis]{pal2010}.
\dm{Recently, \cite{bancelin2012} published the relations extended with
relativistic effects and some non-gravitational forces. It should be noted
that Lie-integration does not handle regularization, i.e., equations
are integrated in proper time by default. However, the method itself
could be applied for regularized forms of the perturbed two-body
problem \citep[see e.g.][for a review about recent methods]{bau2013}.
Due to its properties and implementation techniques, close encounters 
can be handled easily with Lie-series \citep[see also][]{funk2013}.}

The aim of this paper is to present the recurrence relations for
the osculating orbital elements and the mean longitude 
in the case of the planar $N$-body problem. Here we employ a reference frame 
where one of the bodies (i.e., the central body) has been fixed. Choosing 
this reference frame has the advantage that all of the bodies 
orbiting the center 
have constant osculating orbital elements \emph{if} we neglect
mutual interactions. As we show later on, all of the non-trivial
terms depend purely on the mutual terms between the orbiting
bodies. \dm{In other words, trivial cases yield constantly zero series for
the Lie-coefficients. In Sec.~\ref{sec:lienbody} we summarize 
the relations for the fixed-center reference frame, following 
the notations of \cite{hanslmeier1984} and \cite{pal2007}. }
The recurrence equations for constants of motion are derived in 
Sec.~\ref{sec:lieorbitals} while the relations for the mean longitude are
obtained in Sec.~\ref{sec:liemeanlongitude}. Our results and conclusions are
summarized in Sec.~\ref{sec:summary}.

\section{Notations and Lie-series for the N-body problem}
\label{sec:lienbody}

Throughout this paper we follow the conventions used in \cite{hanslmeier1984} 
or \cite{pal2007}. The Newtonian gravitational constant is denoted by $G$,
the mass of the central body is $M$ while the orbiting ones have a mass
of $m_i$ ($1\le i \le N$, hence we deal with $1+N$ bodies). Coordinates
and velocities (with respect to the central body) are denoted 
by $\mathbf{r}_i\equiv r_{ik}$ and $\mathbf{u}_i=u_{ik}$ (where 
$k=1$ or $2$) if we consider vector notations. The components
of these vectors are denoted by $\mathbf{r}_i\equiv(x_i,y_i)$
and $\mathbf{u}_i\equiv(v_i,w_i)$. For simplicity, specific mass is denoted 
by $\mu_i\equiv G(M+m_i)$.

Based on \cite{pal2007}, the relations for the fixed-center problem
are the following series of equations. These are
\begin{equation}
L^{n+1}\mathbf{r}_i = L^{n}\mathbf{u}_i, \label{eq:lbcoord}
\end{equation}
for the coordinates,
\begin{eqnarray}
L^{n+1}\mathbf{u}_i & = &  -\mu_i\sum\limits_{k=0}^{n}\binom{n}{k}L^k\phi_i L^{n-k}\mathbf{r}_i- \label{eq:lbveloc}\\
		    &   &  -G\sum\limits_{j\ne i}m_j \sum\limits_{k=0}^{n}\binom{n}{k}\left[L^k\phi_{ij}L^{n-k}(\mathbf{r}_i-\mathbf{r}_j)+L^k\phi_jL^{n-k}\mathbf{r}_j\right], \nonumber
\end{eqnarray}
for the velocities, 
\begin{eqnarray}
L^n\Lambda_i &=& \sum\limits_{k=0}^{n}\binom{n}{k}L^k\mathbf{r}_iL^{n-k}\mathbf{u}_i, \label{eq:lblambdai}\\
L^n\Lambda_{ij} &=& \sum\limits_{k=0}^{n}\binom{n}{k}L^k(\mathbf{r}_i-\mathbf{r}_j)L^{n-k}(\mathbf{u}_i-\mathbf{u}_j), \label{eq:lblambdaij}
\end{eqnarray}
for the auxiliary quantities $\Lambda_i=\mathbf{r}_i\mathbf{u}_i$ and
$\Lambda_{ij}=(\mathbf{r}_i-\mathbf{r}_j)(\mathbf{u}_i-\mathbf{u}_j)$, and
\begin{eqnarray}
L^{n+1}\phi_i & = & \rho_i^{-2}\sum\limits_{k=0}^{n}F_{nk}^{(-3)}L^{n-k}\phi_i L^k\Lambda_i, \label{eq:lbphii}\\
L^{n+1}\phi_{ij} & = & \rho_{ij}^{-2}\sum\limits_{k=0}^{n}F_{nk}^{(-3)}L^{n-k}\phi_{ij} L^k\Lambda_{ij}. \label{eq:lbphiij}
\end{eqnarray}
for the distances $\rho_i=|\mathbf{r}_i|$, the mutual distances
$\rho_{ij}=|\mathbf{r}_i-\mathbf{r}_j|$ and the reciprocal cubic
distances $\phi_i\equiv\rho_i^{-3}$, $\phi_{ij}\equiv\rho_{ij}^{-3}$. Here
\begin{equation}
F_{nk}^{(-3)}=-3\binom{n}{k}-2\binom{n}{k+1}.
\end{equation}
If we evaluate the above relations in the 
order of equations (\ref{eq:lbcoord}) -- (\ref{eq:lbphiij}), for all values
of $1\le i\le N$ and then increase $n$ by one in each step (thus
starting over with $i=1$, etc.), we obtain the Lie-terms for 
the coordinates and the velocities. The solution of the
original ODE after $\Delta t$ time can be approximated as
\begin{eqnarray}
\mathbf{r}_i(t+\Delta t) & \approx & \sum\limits_{n=0}^{n_{\rm max}}\frac{\Delta t^n}{n!}L^{n}\mathbf{r}_i(t), \label{eq:fsumr} \\
\mathbf{u}_i(t+\Delta t) & \approx & \sum\limits_{n=0}^{n_{\rm max}}\frac{\Delta t^n}{n!}L^{n}\mathbf{u}_i(t). \label{eq:fsumu} 
\end{eqnarray}
Note that for the last value of $n=n_{\rm max}$,
we need only to evaluate equations (\ref{eq:lbcoord}) and (\ref{eq:lbveloc}).
In order to bootstrap these relations, one could consider the fact
that for any quantity $Q$, $L^0Q\equiv Q$. Hence, the above definitions 
and relations for $\Lambda_i$ and $\Lambda_{ij}$ are self-explanatory.

\dm{In the following, we derive the relations for the integrals of motion,
the orbital elements and the mean longitude.}

\section{Relations for the orbital elements}
\label{sec:lieorbitals}

In order to introduce the features of the Lie-series for the
classical Keplerian orbital elements, first, we compute the relations
for the specific angular momentum,
\begin{equation}
C_i=\mathbf{r}_i \wedge \mathbf{u}_i = x_i\dot y_i - y_i\dot x_i = x_iw_i-y_iv_i.
\end{equation}
Since the definition of $C_i$ is similar to the 
relations for $\Lambda_i$ (both are second-order \emph{and} bilinear
functions of the coordinates and velocities), one could expect a similar type 
of relations like equation (\ref{eq:lblambdai}). Indeed, the relations
for the $L^{n}C_i$ terms can be written as 
\begin{equation}
L^{n}C_i=\sum\limits_{k=0}^{n}\binom{n}{k}L^k\mathbf{r}_i\wedge L^{n-k}\mathbf{u}_i= 
	 \sum\limits_{k=0}^{n}\binom{n}{k}\left[L^kx_iL^{n-k}w_i-L^ky_iL^{n-k}v_i\right]. \label{eq:lbci}
\end{equation}
Here, equations for the coordinates and velocities should be computed using
equations (\ref{eq:lbcoord}) -- (\ref{eq:lbphiij}) up to some order of $n\le n_{\rm max}$.
In the case of $N=1$, $L^{n}C_i$ must be equal to $0$ for any $1\le n$
since $C_i\equiv C_1$ is an integral of motion. However, 
equation (\ref{eq:lbci}) does not imply this property. In order to
obtain the values for $L^{n}C_i$, first we compute $L^1C_i$:
\begin{equation}
L^1C_i=LC_i=L(x_iw_i-y_iv_i)=(Lx_i)w_i+x_iLw_i-(Ly_i)v_i-y_iLv_i.
\end{equation}
Since $Lx_i=v_i$ and $Ly_i=w_i$, we get
\begin{equation}
LC_i=v_iw_i+x_iLw_i-w_iv_i-y_iLv_i=x_iLw_i-y_iLv_i.
\end{equation}
Now, equation (\ref{eq:lbveloc}) is substituted for $n=1$:
\begin{eqnarray}
LC_i & = & +x_i\left[-\mu_i\phi_iy_i-G\sum_{i\ne j}m_j[\phi_{ij}(y_i-y_j)+\phi_jy_j]\right]- \nonumber \\
 & & -y_i\left[-\mu_i\phi_ix_i-G\sum_{i\ne j}m_j[\phi_{ij}(x_i-x_j)+\phi_jx_j]\right]. 
\end{eqnarray}
By expanding the above summations and multiplications, the following
can easily be seen. In addition to the Keplerian terms (the first 
ones, proportional to $\mu_i\phi_i$), one part of the 
terms corresponding to the direct perturbations also cancels. Therefore,
\begin{equation}
LC_i = G\sum_{i\ne j}m_j(\phi_{ij}-\phi_j)(x_iy_j-x_jy_i).\label{eq:liecperturb}
\end{equation}
For higher orders, the set of relations can be written as 
\begin{eqnarray}
L^{n}S_{ij} & = & \sum\limits_{k=0}^n\binom{n}{k}(L^kx_iL^{n-k}y_j-L^kx_jL^{n-k}y_i), \\
L^{n+1}C_i  & = & G\sum_{i\ne j}m_j\sum\limits_{k=0}^n\binom{n}{k}L^k\hat\phi_{ij}L^{n-k}S_{ij},\label{eq:lcchigher}
\end{eqnarray}
where we introduce $S_{ij}=x_iy_j-x_jy_i$ and $\hat\phi_{ij}=\phi_{ij}-\phi_j$ 
for simplicity. 

\subsection{Eccentricity and longitude of pericenter}

\dm{In the following, we compute the recurrence relations for the Lagrangian
orbital elements $k=e\cos\varpi$ and $h=e\sin\varpi$.
These are widely used as an equivalent alternative in astrodynamics studies
instead of eccentricity, $e$ and longitude of pericenter, $\varpi$.
In the planar case, $k$ and $h$ are the components of the 
Laplace-Runge-Lenz vector:}
\begin{equation}
\binom{k_i}{h_i}=\frac{C_i}{\mu_i}\binom{+w_i}{-v_i}-\frac{1}{\rho_i}\binom{x_i}{y_i}. \label{eq:khdef}
\end{equation}
Due to the properties of the Lie-operator (linearity and Leibniz' product rule),
the components of the above equation can easily be expanded once
$L\rho_i^{-1}$ is known. Indeed, similarly to $\phi_i=\rho_i^{-3}$, 
it can be shown that 
\begin{equation}
L\rho_i^{-1}=L\left[(\rho_i^2)^{-1/2}\right]=
(-1/2)(\rho_i^2)^{-3/2}L(\rho_i^2)=-1/2\phi_i2\Lambda_i=-\phi_i\Lambda_i,\label{eq:lierhoinverse}
\end{equation}
see also \cite{hanslmeier1984} or \cite{pal2007}. \dm{Now, our goal is to obtain
a relation for $k_i$ and $h_i$ like equation (\ref{eq:liecperturb}) that 
contains only mutual terms. Right after multiplying equation (\ref{eq:khdef}) 
by $\mu_i$, we got the relation}
\begin{equation}
\mu_iLk_i=(LC_i)w_i+C_iLw_i-\mu_i\rho_i^{-1}Lx_i-\mu_iL(\rho_i^{-1})x_i.\label{eq:mulki}
\end{equation}
Then, we have to substitute equations (\ref{eq:liecperturb}), (\ref{eq:lbveloc}),
(\ref{eq:lierhoinverse}), $w_i$ and $\phi_i(x_i^2+y_i^2)$ 
for $LC_i$, $Lw_i$, $L(\rho_i^{-1})$, $Lx_i$ and $\rho_i^{-1}$,
respectively, and then perform a full expansion on equation (\ref{eq:mulki}).
The Keplerian terms indeed cancel and the remaining parts can be written as
\begin{equation}
\mu_iLk_i=G\sum\limits_{i\ne j}m_j\left[\hat\phi_{ij}(w_iS_{ij}+C_iy_j)-C_iy_i\phi_{ij}\right]
\end{equation}
$Lh_i$ can be computed in a similar manner, thus the 
relations for $L(k_i,h_i)$ are
\begin{equation}
L\binom{k_i}{h_i}=\sum\limits_{i\ne j}\frac{Gm_j}{\mu_i}\left[\hat\phi_{ij}\binom{+w_iS_{ij}+C_iy_j}{-v_iS_{ij}-C_ix_j}-C_i\phi_{ij}\binom{+y_i}{-x_i}\right].
\end{equation}
In order to obtain higher order Lie-derivatives, $L^{n+1}(k_i,h_i)$, we should
use Leibniz' product rule for the multilinear expressions appearing in the
above relation. This can either be done directly using the multilinear form
\begin{equation}
L^n(Q_1Q_2\dots Q_m)=\sum\limits_{k_1+k_2+\dots+k_m=n}\frac{n!}{k_1!k_2!\dots k_m!}L^{k_1}Q_1L^{k_2}Q_2\dots L^{k_m}Q_m
\end{equation}
or by introducing auxiliary quantities (e.g. $C_iy_j$, $w_iS_{ij}$)
and subsequently apply the bilinear Leibniz' product rule for these ones.

\subsection{Specific energy and semimajor axis}

The specific energy is defined as 
\begin{equation}
\varepsilon_i=\frac{U_i^2}{2}-\frac{\mu_i}{\rho_i},
\end{equation}
where $U_i=|\mathbf{u}_i|=\sqrt{v_i^2+w_i^2}$.
The semimajor axis can then be computed as $a_i=-\mu_i/(2\varepsilon_i)$.
For simplicity, in the following we compute relations for the
quantity $H_i:=-2\varepsilon_i=\mu_i/a_i$. Using the relations for
$\rho_i^{-1}$ and the velocities (see equation \ref{eq:lbveloc}), 
derivation schemes presented above yields
\begin{equation}
LH_i=2\sum\limits_{i\ne j}Gm_j\left[\phi_{ij}\Lambda_i-\hat\phi_{ij}\hat\Lambda_{ji}\right],\label{eq:lieenergy}
\end{equation}
where we introduce $\hat\Lambda_{ji}=x_jv_i+y_jw_i$.
The higher order Lie-derivatives are then obtained as it is described
at the end of the previous section.

\section{Relations for the mean longitude}
\label{sec:liemeanlongitude}

The previously obtained relations for the orbital elements 
can applied not only for closed (circular or elliptic) orbits but for parabolic
and hyperbolic orbits, as well. In the following, due to its relevance,
we handle only closed orbits. Hence, eccentricity $e=\sqrt{k^2+h^2}$ 
is expected to be smaller than unity for all orbits and
the reciprocal semimajor axis $\mu/a=-2\varepsilon=H$ is also positive. 

\dm{The mean longitude is the only related quantity which is defined 
for both circular and elliptical orbits \emph{and} which is an 
analytic function of the coordinates and 
velocities \citep[see e.g.][]{pal2009}. Therefore, in the following we 
ignore the eccentric, mean and true anomalies from the computations.
It should be noted that some quantities like $e\sin E$ or $e\cos E$
also behaves analytically in the $e\to 0$ limit, hence Lie-series
can also be defined for these \citep[where $E$ denotes the
eccentric anomaly, see e.g.][]{pal2009}.}

\subsection{Full expansion of the mean longitude}

The mean longitude $\lambda_i$ can be computed using 
the analytic equation
\begin{equation}
\lambda_i=\arg\left[+\hat\rho_i w_i+h_i\Lambda_i,-\hat\rho_i v_i-k_i\Lambda_i\right]-\frac{\Lambda_i}{C_i}J_i.
\end{equation}
Here we introduced $J_i=\sqrt{1-e_i^2}=b_i/a_i$, the oblateness of the orbit and
$\hat\rho_i=\rho_i(1+J_i)$. Regarding to the differentiation, 
the $\arg(x,y)$ function behaves like the 
arc tangent function, $\mathrm{arc\,tg}(y/x)$: 
\begin{equation}
\mathrm{d}\left[\arg(x,y)\right]=\mathrm{d}\left[\mathrm{arc\,tg}\left(\frac{y}{x}\right)\right]=
\frac{x\mathrm{d}y-y\mathrm{d}x}{x^2+y^2}.
\end{equation}
The first-order Lie-derivative of $\lambda_i$ is then
\begin{equation}
L\lambda_i=\frac{(\hat\rho_i v_i+k_i\Lambda_i)L(\hat\rho_i w_i+h_i\Lambda_i)-(\hat\rho_i w_i+h_i\Lambda_i)L(\hat\rho_i v_i+k_i\Lambda_i)}{(\hat\rho_i w_i+h_i\Lambda_i)^2+(\hat\rho_i v_i+k_i\Lambda_i)^2}-L\left(\frac{\Lambda_i}{C_i}J_i\right)\label{eq:llambdai}.
\end{equation}
The denominator of the first (apparently large) fraction can significantly
be simplified to the form $(1+J_i)^2C_i^2$. Now one has to simplify
the above equation in order to depend mostly on the mutual interactions.
Since $L\lambda_i=\dot\lambda_i=n_i\ne 0$ even for non-perturbed orbits,
this simplification cannot \dm{be homogeneous with respect to $Gm_j$. 
In the following, we deal with the perturbed and non-perturbed terms 
separately and expand the above equation into two parts. 
The expansion of the numerator in the first fraction of 
equation (\ref{eq:llambdai}) yields}
\begin{eqnarray}
& (\hat\rho_i v_i+k_i\Lambda_i)L(\hat\rho_i w_i+h_i\Lambda_i)-(\hat\rho_i w_i+h_i\Lambda_i)L(\hat\rho_i v_i+k_i\Lambda_i) = \label{eq:lambdaexpand} \\
& = \hat\rho_i^2(v_iLw_i-w_iLv_i)+(\Lambda_iL\hat\rho_i-\hat\rho_iL\Lambda_i)(w_ik_i-v_ih_i)+ \nonumber \\
& + \hat\rho_i\Lambda_i(v_iLh_i-w_iLk_i+k_iLw_i-h_iLv_i)+\Lambda_i^2(k_iLh_i-h_iLk_i). \nonumber
\end{eqnarray}
The terms appearing above can be expanded as:
\begin{eqnarray}
v_iLw_i-w_iLv_i & = & \mu_i\phi_iC_i+G\sum\limits_{i\ne j}m_j\left[\phi_{ij}C_i-\hat\phi_{ij}\hat C_{ji}\right],\\
v_ih_i-w_ik_i & = & C_i\left(\frac{1}{\rho_i}-\frac{U_i^2}{\mu_i}\right), \\
v_iLh_i-w_iLk_i+k_iLw_i-h_iLv_i & = & -C_i\phi_i\Lambda_i- \sum\limits_{i\ne j}Gm_j\hat\phi_{ij}S_{ij}\left(\frac{1}{\rho_i}+\frac{U_i^2}{\mu_i}\right), \\
k_iLh_i-h_iLk_i & = & \sum\limits_{i\ne j}\frac{Gm_j}{\mu_i}\left[-\frac{C_i^2}{\mu_i}\hat\phi_{ij}\hat C_{ji}+\frac{C_i^3}{\mu_i}\phi_{ij}+\right. \nonumber \\
	& & +\left.\frac{\hat\phi_{ij}}{\rho_i}(\Lambda_iS_{ij}+C_iR_{ij})-C_i\rho_i\phi_{ij}\right], \\
\Lambda_iL\hat\rho_i-\rho_iL\Lambda_i & = & (1+J_i)(\Lambda_i^2\rho_i^{-1}-\rho_iL\Lambda_i)+\Lambda_i\rho_iLJ_i
\end{eqnarray}
and
\begin{eqnarray}
L\Lambda_i & = & \left(U_i^2-\frac{\mu_i}{\rho_i}\right)+\sum\limits_{i\ne j}Gm_j\left[\hat\phi_{ij}R_{ij}-\phi_{ij}\rho_i^2\right].
\end{eqnarray}
where $\hat C_{ji}=x_jw_i-y_jv_i$ and $R_{ij}=\mathbf{r}_i\mathbf{r}_j=x_ix_j+y_iy_j$.

Using the well-known relations from classical celestial mechanics,
it can be shown that the double-negative specific energy, $H_i$
relates to the oblateness $J_i$ and the specific angular momentum $C_i$ 
as $C_i^2H_i=J_i^2\mu_i^2$. From this relation, by taking the Lie-derivative
of both sides, we got 
\begin{equation}
LJ_i=J_i\left(\frac{LC_i}{C_i}+\frac{LH_i}{2H_i}\right).
\end{equation}
Therefore, the last term in equation (\ref{eq:llambdai}) can be written as
\begin{eqnarray}
L\left(\frac{\Lambda_i}{C_i}J_i\right) & = & -\frac{LC_i}{C_i^2}J_i\Lambda_i+\frac{J_i}{C_i}L\Lambda_i+\frac{\Lambda_i}{C_i}LJ_i= \nonumber \\
 & = & -\frac{LC_i}{C_i^2}J_i\Lambda_i+\frac{J_i}{C_i}L\Lambda_i+\frac{\Lambda_i}{C_i}\frac{J_i}{C_i}LC_i+\frac{\Lambda_i}{C_i}\frac{J_i}{2H_i}LH_i.
\end{eqnarray}
Here the first and third terms cancel each other, thus
\begin{equation}
L\left(\frac{\Lambda_i}{C_i}J_i\right)  = \frac{J_i}{C_i}L\Lambda_i+\frac{J_i\Lambda_i}{2C_iH_i}LH_i.\label{eq:ljcexpand}
\end{equation}

\subsection{The non-perturbed part}

From the above series of equations we collect those where
terms after the summation $\sum\limits_{i\ne j}Gm_j(\cdot)$ do not 
occur. This part, denoted as $L\lambda_i|_0$ is 
\begin{eqnarray}
L\lambda_i|_0 & = & -\frac{J_i}{C_i}\left(U_i^2-\frac{\mu_i}{\rho_i}\right)+\frac{1}{C_i^2(1+J_i)^2}\bigg\{\hat\rho_i^2\mu_i\phi_iC_i-\hat\rho_i\Lambda_i^2C_i\phi_i-\bigg. \nonumber \\
	& & \bigg.-\left[\Lambda_i^2\rho_i^{-1}(1+J_i)-\hat\rho_i\left(U_i^2-\frac{\mu_i}{\rho_i}\right)\right]C_i\left(\frac{1}{\rho_i}-\frac{U_i^2}{\mu_i}\right)\bigg\}. \label{eq:llambda0}
\end{eqnarray}
By substituting the relations $U_i^2-\mu_i/\rho_i=\mu_i/\rho_i-H_i$, 
$C_i^2H_i=J_i^2\mu_i^2$ and $\Lambda_i^2+C_i^2=U_i^2\rho_i^2$,
equation (\ref{eq:llambda0}) \dm{can greatly be simplified to
obtain Kepler's Third Law:}
\begin{equation}
L\lambda_i|_0 = \frac{\mu_i^2J_i^3}{C_i^3} = 
\frac{1}{\mu_i}H_i^{3/2}=\sqrt{\frac{\mu_i}{a_i^3}}.
\end{equation}

\subsection{The perturbed part}

Let us write the full Lie-derivative of $L\lambda_i$ in the form
\begin{equation}
L\lambda_i=\frac{1}{\mu_i}H_i^{3/2}+\sum\limits_{i\ne j}Gm_j [L\lambda]_{ij}.
\end{equation}
This is similar to the forms obtained for the angular momentum, specific
energy and Lagrangian orbital elements, with the exception of the presence
of the term related to Kepler's Third Law. The goal now is to compute
the terms $[L\lambda]_{ij}$ as simple as possible. It can be shown that
this term is
\begin{eqnarray}
[L\lambda]_{ij} & = & +\frac{\hat\phi_{ij}}{1+J_i}\left[\left(-\frac{2J_i(1+J_i)}{C_i}+\frac{2C_i}{\mu_i\rho_i}\right)R_{ij}-\left(\frac{\rho_i}{\mu_i}+\frac{C_i^2}{\mu_i^2}\right)\hat C_{ji}\right] + \nonumber \\
& & +\frac{\phi_{ij}}{1+J_i}\left[\frac{C_i^3}{\mu_i^2}-\frac{C_i}{\mu_i}\rho_i+\frac{2J_i(1+J_i)}{C_i}\rho_i^2\right]. \label{eq:llambdaij}
\end{eqnarray}
\dm{The deduction of the above equation has the following steps.
First, one should fully expand equation (\ref{eq:lambdaexpand}) while 
keeping only the terms $\sum Gm_j(\cdot)$. 
Then, it is divided by $(1+J_i)^2C_i^2$ after which
we add the expansion of equation (\ref{eq:ljcexpand}), still keeping 
only the terms $\sum Gm_j(\cdot)$.}
This equation (\ref{eq:llambdaij}) can be simplified in terms of computation 
implementation by introducing the dimensionless quantity $g_i=\mu_i\rho_iC_i^{-2}$:
\begin{eqnarray}
[L\lambda]_{ij} & = & +\frac{\hat\phi_{ij}}{1+J_i}\left[\frac{2}{C_i}\left(g_i^{-1}-J_i(1+J_i)\right)R_{ij}-\frac{\rho_i}{\mu_i}\left(g_i^{-1}+1\right)\hat C_{ji}\right] + \nonumber \\
& & +\frac{\phi_{ij}}{1+J_i}\frac{\rho_i^2}{C_i}\left[g_i^{-2}-g_i^{-1}+2J_i(1+J_i)\right]. \label{eq:llambdaij2}
\end{eqnarray}
Therefore, the first Lie-derivative of $\lambda_i$ can be written as
\begin{eqnarray}
L\lambda_i & = & \frac{1}{\mu_i}H_i^{3/2}+\sum\limits_{i\ne j}Gm_j\left[\hat\phi_{ij}(A_{\mathrm R}R_{ij}+A_{\mathrm C}\hat C_{ji})+\phi_{ij}A_0R_{ii}\right]
\end{eqnarray}
where 
\begin{eqnarray}
A_{\mathrm R}	& = & \frac{2}{C_i}\left(\frac{g_i^{-1}}{1+J_i}-J_i\right), \label{eq:lambdaar}\\
A_{\mathrm C}	& = & \frac{\rho_i}{\mu_i}\left(\frac{1+g_i^{-1}}{1+J_i}\right),  \text{and} \label{eq:lambdaac}\\
A_0	  	& = & \frac{1}{C_i}\left(\frac{g_i^{-2}-g_i^{-1}}{1+J_i}+2J_i\right). \label{eq:lambdaa0}
\end{eqnarray}
Higher order derivatives can then be computed using the relation
\begin{eqnarray}
L^{n+1}\lambda_i & = & \frac{1}{\mu_i}L^n\left(H_i^{3/2}\right)=+\sum\limits_{i\ne j}Gm_j\sum\limits_{k+p+q=n}\frac{n!}{k!p!q!}\times \nonumber \\
& & \times \left[L^k\hat\phi_{ij}\left(L^pA_{\mathrm R}L^qR_{ij}+L^pA_{\mathrm C}L^q\hat C_{ji}\right)+L^k\phi_{ij}L^pA_0L^qR_{ii}\right]
\end{eqnarray}
Let us suppose that the Lie-derivatives of the arbitrary quantity $Q$ are known 
up to the order of $n+1$. It can be shown by mathematical induction
that the $(n+1)$th Lie-derivative of $Q^p$ can be computed using the relation
\begin{equation}
L^{n+1}Q^p=Q^{-1}\sum\limits_{k=0}^n\left[p\binom{n}{k}-\binom{n}{k+1}\right]L^{n-k}\left(Q^p\right)L^{k+1}Q\label{eq:liepower}
\end{equation}
\dm{By substituting $p=3/2$, this relation can be 
used to compute $L^nH_i^{3/2}$ if higher order derivatives of $H_i$ are known. 
In addition, equation (\ref{eq:liepower}) can be exploited in order to 
compute $(1+J_i)^{-1}$, $C_i^{-1}$, $C_i^2$ and $g_i^{-2}$. 
The additional terms $A_{\mathrm R}$, $A_{\mathrm C}$
and $A_0$ depend only on the $i$th orbit. Hence, the relatively complex
equations (\ref{eq:lambdaar}) -- (\ref{eq:lambdaa0}) 
are only computed $N$ times in a single iteration, 
instead of $N^2/2$. Therefore, these calculations do not 
significantly increase the total computing time  for larger number of bodies.}

\section{Conclusions and summary}
\label{sec:summary}

In this paper we presented recurrence formulae of the orbital elements 
related to the planar $N$-body problem. As we showed, the structure 
of these formulae depends only on the terms related to the 
mutual interactions. Therefore, the relations for the two-body problem
reduces to a constant motion that can be integrated with arbitrary step size.
\dm{It should be noted that although the presented procedure still requires
the computation of higher order derivatives of coordinates and velocities,
these relations are exploited as auxiliary equations for computing
the mutual terms and these are not integrated directly. }

\dm{In order to estimate the merits of using the orbital elements instead of the 
coordinates and velocities, we can compare, for instance, the magnitude of the
terms $L^kC_i$ when these are computed using equation (\ref{eq:lbci})
or equation (\ref{eq:lcchigher}). In the unperturbed case, the
latter one yields exactly zero while roundoff errors initiate an exponential
growth in the higher order derivatives yielded by naive computation. 
Using double-precision arithmetic and bootstrapping with unity specific 
mass and angular momentum, the roundoff errors accumulate to unity around 
the order of $k\approx 19\dots 21$, depending on the initial eccentricity 
and orbital phase. In addition, for a given step size and desired precision,
employing orbital elements instead of coordinate components 
decrease the integration order $n_{\rm max}$. For weakly 
perturbed systems (like the inner Solar System), 
this decrement can be a factor of $\sim 2$. This would naively yield 
a gain of $\sim 4$ in computing time due its $\mathcal{O}(n_{\rm max}^2)$ 
dependence. However, the additional computations needed by the orbital
elements make a practical implementation less efficient. 
Our initial analysis also showed that the higher the perturbations, 
the less the gain in the integration order. 
In the case of the outer Solar System (where $m_i/M \lesssim 10^{-3}$), 
this gain in the decrease of the maximum of derivative order 
is less prominent.}

Following studies could investigate the relations for the spatial problem.
In some cases, this extension could be straightforward for
scalar quantities like the specific energy. \dm{Care must be taken in the cases 
where pseudo-scalars (like $C_i$) or explicit coordinates occur.}
Another interesting point can be the elimination of the need for computing
the recurrence formulae for coordinates and velocities and employ 
directly the orbital elements. 

\vspace*{1ex}

\noindent
\textbf{Acknowledgments.}
The author would like to thank the anonymous referees for their
valuable comments. The author also thanks L\'aszl\'o Szabados 
for the careful proofreading.
This work has been supported by the Hungarian Academy of
Sciences via the grant LP2012-31. 

{}

\end{document}